\documentclass[twocolumn,aps,showpacs]{revtex4}
\usepackage{graphics}
\usepackage{hyperref}
\begin{document}

\title{Directed avalanche processes with underlying 
interface dynamics}

\author{Chun-Chung Chen}

\author{Marcel \surname{den Nijs}}

\affiliation{Department of Physics, University of Washington,
Seattle, WA 98195, USA}

\date{February 18, 2002}

\begin{abstract}
We describe a directed avalanche model; a slowly unloading sandbox
driven by lowering a retaining wall.  The directness of the dynamics
allows us to interpret the stable sand surfaces as world sheets of
fluctuating interfaces in one lower dimension.  In our specific case,
the interface growth dynamics belongs to the Kardar-Parisi-Zhang (KPZ)
universality class.  We formulate relations between the critical
exponents of the various avalanche distributions and those of the
roughness of the growing interface.  The nonlinear nature of the
underlying KPZ dynamics provides a nontrivial test of such generic
exponent relations.  The numerical values of the avalanche exponents
are close to the conventional KPZ values, but differ sufficiently to
warrant a detailed study of whether avalanche correlated Monte Carlo
sampling changes the scaling exponents of KPZ interfaces.  We
demonstrate that the exponents remain unchanged, but that the traces
left on the surface by previous avalanches give rise to unusually
strong finite-size corrections to scaling. This type of slow
convergence seems intrinsic to avalanche dynamics.
\end{abstract}


\maketitle

\section{Introduction}

Avalanche phenomena are common in nature.  Examples range from
accumulating snow on mountain slopes, slow shearing between
continental plates~\cite{Carlson1989}, rerouting in river networks, to
creeping magnetic flux lines in super conductors~\cite{Bassler1998}.
Following the work by Bak {\it et al.}~\cite{Bak}, physicists aim to
capture the essential aspects of such dynamical systems with simple
automaton processes, commonly referred to as sandpile models and
self-organized criticality (SOC).  Impressive successes have been
achieved, like reproducing power-law distributions in avalanche events
similar to those observed in nature, and the start of a classification
scheme of such processes in terms of so-called universality
classes~\cite{Ben-Hur1996}.  Unfortunately most of these are numerical
in nature.  Analytical exact results remain rare.

Directed avalanche phenomena form a subclass of these SOC processes.
Dhar and Ramaswamy introduced the first directed sandpile model and
solved it exactly~\cite{Dhar1989}.  This was possible because in their
model the avalanche propagation is governed solely by its two edges,
and those two follow independent random walk dynamics.  Tadi\'c and
Dhar~\cite{Tadic1997} introduced a directed model in which particles
are allowed to pile up beyond the critical height, by replacing the
automaton's deterministic toppling rule by a stochastic
one~\cite{Tadic1997}.  The density of critical sites tunes itself and
at distances far from the driving edge the propagation of active sites
approaches the directed percolation (DP)~\cite{Kinzel1983} threshold.
The scaling properties of the avalanche distributions are thus linked
to the critical exponents characterizing the DP universality class.
Another example of a stochastic directed avalanche process is the
model introduced and studied numerically by Pastor-Satorras and
Vespignani~\cite{Pastor-Satorras2000}.  Similar as in the above model
by Dhar and Ramaswamy, the stable landscape configurations (between
avalanche events) lack internal correlations in the stationary state.
This allowed Paczuski and Bassler~\cite{Paczuski2000} and also Kloster
{\it et al.}~\cite{Kloster2001} to link this dynamic process to
so-called Edwards-Wilkinson~\cite{Edwards1982} (EW) interface growth
and to derive the exact scaling exponents of the avalanche
distributions.

This novel world sheet type connection between avalanche dynamics and
interface growth is particularly promising, because interface dynamic
processes like EW and Kardar-Parisi-Zhang~\cite{Kardar1986} (KPZ)
growth are very well understood, in particular in 1+1 dimensions
(1+1D) where the scaling properties are known exactly.  However, the
above models that are linked to EW type growth are rather poor
examples, because EW growth is described by a simple linear stochastic
(diffusion type) Langevin equation; correlations factorize, and
important caveats in the relation to avalanche dynamics can be
obscured by this simplicity.

We set out to generalize this approach to nonlinear interface dynamic
processes, and recently introduced a directed unloading sandbox
model~\cite{Chen2002} in which the two dimensional (2D) avalanche
dynamics relates to 1+1D KPZ type interface growth.  We derived
exponent relations between the avalanche and interface growth scaling
properties, which are generic, and valid beyond our specific model.
Our numerical results for the avalanche distributions (for length,
width, depth, and mass) follow indeed these exponent relations.
Moreover, the avalanche critical exponents obey the predicted KPZ
values within a few percent, an accuracy typical to avalanche
simulations.  However, our numerical accuracy is better than that;
mostly because of a careful finite-size scaling (FSS) analysis.  The
exponents seem to converge to values that are slightly different from
the KPZ values.

This left us with a puzzle.  What is the origin of these small
deviations?  Is this a fundamental effect; or do the exponents
ultimately converge to the KPZ values, but with unusually large
corrections to scaling.  In this paper we address these issues.  We
also provide a more detailed discussion of these world-sheet-type
relationships between avalanche and interface growth dynamics.  Our
first paper was short and did not include many of the details that are
crucial for the analysis presented here.

The fundamental difference between conventional KPZ interface growth
and avalanche dynamics arises from the averaging process over KPZ type
space-time world sheets.  In normal Monte Carlo (MC) simulations of
interface growth the distribution functions are determined in terms of
ensemble averages over a set of totally uncorrelated space-time MC
runs.  In contrast, the avalanche dynamics gives rise to KPZ world
sheets that are strongly correlated.  Two subsequent MC runs are
identical except inside a single avalanche area.  This difference in
averaging, uncorrelated versus avalanche correlated MC runs, therefore
emerges as a key issue for understanding the scaling properties of
avalanche dynamics.  This issue did not arise in the earlier EW type
avalanche models due to the linear nature of the EW process.  However
for nonlinear dynamics, like KPZ, avalanche-correlated-type sampling
could well lead to novel interface scaling exponents.

Speaking against a shift in the values of the exponents, are arguments
like: the KPZ stationary state, i.e., the sand surface profile far way
from the driving edge, can not be affected by the
avalanche-correlated-type averaging, because large avalanches that
span the entire width of the box occur periodically.  These completely
refresh the surface far way from the driving edge regularly, and thus
wipe out all correlations between MC runs.  This suggests that we are
only dealing with much larger than usual corrections to scaling.  The
details are more complex than this simple argument, but we will
establish that indeed the exponent values do not change.

The paper is organized as follows.  In the next section we present the
unloading sandbox model.  In Sec.~\ref{moti} we comment on how
directed avalanche dynamics can be linked to interface growth in one
lower dimension.  Next, in Sec.~\ref{KPZ}, we show that in the
interface growth interpretation our specific model belongs to the KPZ
universality class.  In Sec.~\ref{scaling} we derive the generic
exponent relations between interface growth and directed avalanche
dynamics, and in Sec.~\ref{num} we test this numerically for our
specific model.

In the second half of this paper we address the small deviations in
the numerical values of the exponents from those of conventional KPZ
growth.  In Sec.~\ref{corrMC} we present numerical results detailing
how the traces left on the surface profile by previous avalanches
influence both the avalanche exponents and the interface growth ones.
These scars in the rough surface enhance the surface roughness.  We
cast this enhanced interface roughness in terms of corrections to
scaling, and determine what value the critical dimension of the
corresponding irrelevant operator $O_{\rm sc}$ (in the sense of
renormalization theory) should have.  Next, we identify the geometric
meaning of $O_{\rm sc}$, starting with a study of the one dimensional
(1D) version of our model where a similar phenomenon takes place,
Sec.~\ref{1Dedge}.  In 1D the interface growth process is a simple
random walk, and the avalanche correlated sampling relates to the
scaling properties of merging random walkers.  $O_{\rm sc}$ represents
the distribution of avalanche endpoints in the 1D surface, and can be
studied directly from the rounding of the surface profile near the
driving edge.  In Sec.~\ref{2Dedge} we return to the full 2D case.
The scars of previous avalanches form lines on the surface.  We
identify $O_{\rm sc}$ with the angle these lines make with respect to
the direction perpendicular to the driving edge, and confirm with an
analytic argument that the critical dimension of $O_{\rm sc}$ is equal
to $x_{\rm sc}=-z$ with $z$ the KPZ dynamic exponent.  Finally, we
summarize our results in Sec.~\ref{sum}.

\section{An Unloading Sandbox}\label{model}

\begin{figure}
{\centering
\resizebox*{0.8\columnwidth}{!}
{\includegraphics{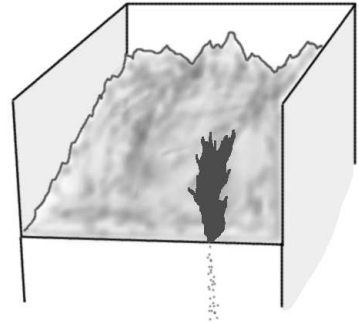}}
\par}
\caption{Sandbox with a slowly lowering retaining wall}
\label{sandbox}
\end{figure}

Imagine a box filled with granular material, as illustrated in
Fig.~\ref{sandbox}.  One of its four retaining walls is slowly
lowered, such that the sand spills out from that side, and thus slowly
unloads the box and establishes a sloped surface.  In the quasistatic
limit, the wall moves slow enough that the unloading events can be
described as distinct avalanches.  The box can be three dimensional,
leading to 2D avalanche dynamics on a 2D surface, or can be 2D (like
in a very narrow box) giving rise to 1D avalanches on a 1D surface.

Inspired by this we consider a so-called solid-on-solid model defined
on a 2D lattice.  Height variables $h({\bf r})$ are defined on a
square lattice.  We will consider two versions of the model.  In the
continuous height version, the heights are real numbers.  In the
discrete model, the heights are integers, $h({\bf r})=0,\pm 1,\pm
2,\cdots$.  The former corresponds to a continuous material without
internal structure, but strong cohesion up to a specific length scale
$s_c$, while the latter corresponds to layered material where the
surface height is quantized.

\begin{figure}
{\centering
\resizebox*{0.8\columnwidth}{!}
{\includegraphics{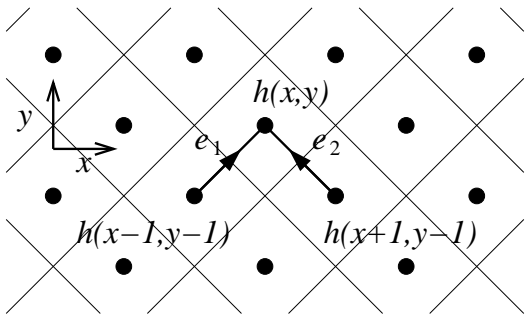}}
\par}
\caption{Lattice structure of sandbox model in 2D}
\label{2D sandbox lattice}
\end{figure}

The 2D lattice is rotated diagonally such that the propagation
direction of the avalanche is along the diagonal direction denoted by
$y$. This is the direction in which the avalanche will run. Throughout
this paper the coordinate perpendicular to $y$ will be denoted by $x$.
Figure~\ref{2D sandbox lattice} illustrates this geometry.

The configurations are subject to the following stability condition.
The column of particles on site ${\bf r}=(x,y)$ is supported by the
two columns, ${\bf r}_l=(x-1,y-1)$ and ${\bf r}_r=(x+1,y-1)$ directly
below it and is stable when its height is less than the minimum of the
heights at these two supporting sites increased by a fixed amount
\begin{equation}
\label{stability}
h({\bf r})\leq \min \left[ h({\bf r}_l),h({\bf r}_r)\right] +s_{c}.
\end{equation}
$s_{c}$ is a constant.  In the version of our model where the heights
are continuous variables $s_{c}$ represents the only length scale in
the $h$ direction and can be set equal to 1 without loss of
generality.  Throughout this paper we will also set $s_{c} = 1$ in the
discrete $h$ model.

Consider a stable configuration, after $\tilde t-1$ avalanches.  The
$\tilde t$-th avalanche is triggered at the highest site ${\bf
r}=({\bf x}_{\tilde t},0)$, on the $y=0$ driving boundary (or, in the
discrete height model, by randomly choosing one of the highest sites)
and reducing its height by a random amount $0<\eta _{\tilde t}\leq
s_{c}$.  This likely creates unstable sites in the next $y=1$
row. Those are updated by replacing their height by an amount equal to
the lowest of the two supporting columns in the previous row and then
adding an uncorrelated random amount $0\leq \eta ({\bf r})\leq s_{c}$
with uniform distribution, as
\begin{equation}
\label{toppling rule}
h({\bf r}) \rightarrow \min \left[ h({\bf r}_l),h({\bf r}_r)\right]
+\eta ({\bf r}).
\end{equation}
This updating continues row by row until all the sites are stable
again.  Only after that the next avalanche is started.  The toppling
of a site only effects the stability of the two sites immediately
above it in the next $y$-row.  Therefore we can update the system
row-by-row in increasing order of $y$.

Direct experimental realizations of this unloading sandbox model are
not our immediate concern (the focus is on establishing a generic
theoretical relationship between avalanche dynamics and interface
growth), but we expect that this model is applicable to actual
experimental unloading sandboxes.  One of the most important issues in
this context is the row-by-row nature of the toppling rule.  This is a
crucial feature for our purposes, allowing the identification with KPZ
interface growth (in the next section).  In real unloading sandboxes
the sand removed from row $y$ rolls down hill and likely disturbs the
already stabilized lower surface levels. Experimental realizations can
avoid this from happening, e.g., by choosing very light grains
(compared to the cohesion forces).  Note that our dynamic rule does
not allow the build-up of any pockets (deeper than $s_c$) on the
surface that might trap such downward rolling grains.

Conservation laws are crucial to avalanche dynamics.  Unlike most
avalanche processes, our model does not conserve mass while the
avalanche propagates.  That might raise the specter of our model not
being (self-organized) critical.  The connection to KPZ growth (an
intrinsic critical process) dispels this phantom.  Moreover, the
global slope of the surface is preserved during each avalanche run,
and conservation of steps in the profile plays the role analogous to
conservation of mass.

The analysis of the dynamics involves distribution functions of
various characteristic features of the avalanches, The common examples
are: length, width, depth, and mass.  The avalanche length $ l $ will
be defined throughout this paper as the maximum distance $ y $ the
avalanche travels from the driving edge; the width $ w $ as the
maximum departure of the $x$-coordinate (perpendicular to the
propagation direction) from the trigger point $x$-coordinate; the
depth $ \delta $ as the maximum height change the avalanche creates at
any of the affected sites; and the mass $ m $ as the total amount of
material removed by the avalanche.\label{definition of l w d m}

\section{Avalanches versus epitaxial interface growth}\label{moti}

The focus of this paper is on how the above avalanche dynamics relates
to interface growth in one lower dimension.  Each stable sloped
surface configuration of a directed sandpile can be reinterpreted as a
world sheet (space-time configuration) of an interface in one lower
spatial dimension. The direction in which the avalanches propagate
plays the role of time and the perpendicular coordinates the role of
space.  Our 2D unloading sandbox is equivalent to a 1D growing
interface.  Such an interpretation makes sense only when the stability
condition and the avalanche dynamic rule is directional and local in
space-time, such that causality is not violated in the interface
growth interpretation.  The stability condition~(\ref{stability}) and
toppling rule~(\ref{toppling rule}) of our model are row-by-row in
nature and therefore indeed Markovian in this sense.

Every stable configuration of the sandpile represents a possible
interface growth life line (space-time-evolution interface world
sheet).  The conventional procedure for determining the scaling
properties of growing interfaces is to average over a large set of
completely independent MC runs.  This would mean, in sandbox
language, an ensemble average over completely refreshed surfaces, each
totally uncorrelated from the previous one (except typically for the
initial condition in row $y=0$).  The toppling rule~(\ref{toppling
rule}) is applied to all sites in every row, and repeated row-by-row,
instead of only the unstable sites created by toppling only the
highest site in the initial row.

In avalanche dynamics, however, two subsequent growing interface life
lines in this ensemble differ only inside the avalanche area.  From
the interface growth perspective this represents a rather peculiar and
dangerous correlated type MC run averaging procedure.  The MC runs of
KPZ space-time configuration are strongly correlated, and this raises
the specter of a change in the interface roughness scaling properties.
The numerical evidence, presented below is sufficiently ambiguous that
this issue will preoccupy us in the second half of this paper.

\section{KPZ growth}\label{KPZ}

In this section we demonstrate that the interface growth model
conjugate to the unloading 2D sandbox belongs to the 1+1D KPZ
universality class.  The time evolution of the interface is governed
by the the toppling rule of the sand model with $ y $ in
Eq.~(\ref{toppling rule}) representing time $ t $,
\begin{equation}
\label{interface dynamic rule}
h(x,t+1)=\min \left[ h(x+1,t),h(x-1,t)\right] +\eta (x,t).
\end{equation}
In the conventional global type interface evolution (i.e., totally
refreshing non-avalanche-type uncorrelated MC runs) every site in row
$t+1$ is updated according to this rule.

\begin{figure}
{\centering
\resizebox*{0.8\columnwidth}{!}
{\includegraphics{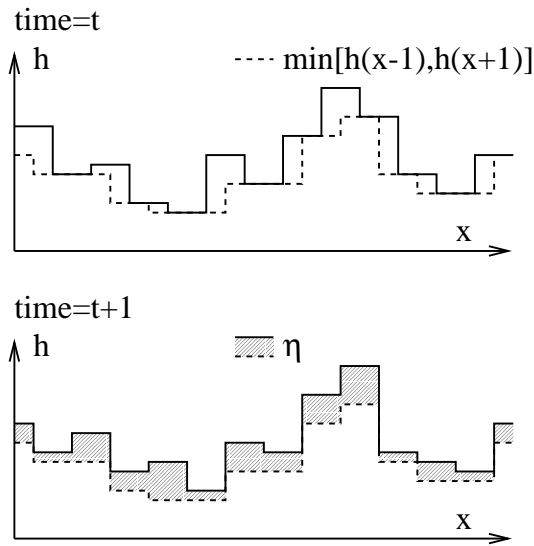}}
\par}
\caption{The interface growth dynamics described by
Eq.~(\ref{interface dynamic rule}) with upper panel showing movement
of steps (from the drawn to dashed line) and lower panel random
depositions (shaded area) to the interface}
\label{interface dynamics}
\end{figure}

Figure~\ref{interface dynamics} illustrates the interface dynamics for
one time step, $ t\rightarrow t+1 $. Conceptually the time step can be
split into two parts; the deterministic $\min[~]$ operator part and
the stochastic random deposition $\eta$ part.

Note that because of the diagonal orientation of the square lattice
(see Fig.~\ref{2D sandbox lattice}), the lattice sites are not
``stationary in time''. The conceptually easiest interpretation to
resolve this flip-flopping is to first double the number of lattice
sites and then to require them to be paired alternately with their
right or left neighbors at even and odd times; at even times sites
$2n$ and $2n+1$ are fused to be at equal heights and at odd times the
$2n-1$ and $2n$ sites.

The upper panel shows the deterministic first half of the update (from
the drawn to the dash line). The partners switch and the $\min[~]$
operation equalizes their heights by choosing the lowest of the two.
so this step always removes material.

This can be interpreted also in terms of a movement of the steps in
the interface. All up-steps move to the right and all down-steps to
the left; while up and down steps merge when they meet at one site.

The lower panel illustrates the second half of the update.  The height
of each fused pair increases by a random amount $ 0\leq \eta \leq
s_{c} $.

Deposition-type interface dynamics like this typically belongs to the
KPZ universality class~\cite{Kardar1986}.  Indeed, Eq.~(\ref{interface
dynamic rule}) can be rewritten as
\begin{eqnarray}
h(x,t+1) & = & \frac{1}{2}\left[ h(x+1,t)+h(x-1,t)\right] \nonumber \\
& - & \frac{1}{2}\left| h(x+1,t)+h(x-1,t)\right| + \eta (x,t),
\end{eqnarray}
and from this easily be identified to be a discrete form of the KPZ
Langevin equation,
\begin{equation}
\label{KPZ-equation}
\frac{\partial h}{\partial t}=\nabla ^{2}h-\frac{\lambda }{2}\left(
\nabla h\right) ^{2}+\eta .
\end{equation}

The crucial point is that the coefficient of the nonlinear term
$\lambda$ is clearly present.  There is no hidden special symmetry of
some kind that makes it vanish by accident.  At $\lambda=0$, the KPZ
equation would reduce to EW growth.

To confirm the KPZ nature and make sure that the $\lambda$ is large
enough that corrections to scaling from the EW point ($\lambda=0$) are
not obscuring the KPZ scaling, we perform MC simulations on the
interface dynamics as illustrated in Fig.~\ref{interface dynamics}.
The MC runs are completely independent.

We measure the time evolution of the interface width $W$ defined as
\begin{equation}
  W^2(L_x,t)\equiv \langle\overline{(h-\bar h)^2}\rangle
\end{equation}
with over bars (angle brackets) indicating average over $x$ (ensemble).
Starting from, e.g., a flat initial condition it should scale as
\begin{equation}
  W\sim t^{\beta }
\label{beta definition}
\end{equation}
at intermediate times $ 0\ll t\ll L_x^{z} $, and saturate at
\begin{equation}
  W\sim L_x^{\alpha }
  \label{alpha definition}
\end{equation}
for $t\gg L_x^{z}$; with $L_x$ the length of the 1D interface.  The
exponents for the KPZ universality class in 1+1D are known exactly
with $\alpha=1/2$, $\beta=1/3$, and $z\equiv\alpha/\beta=3/2$.

The numerical results are shown in Fig.~\ref{interface scaling
exponents}.  The values of $\alpha(L_x)$ are obtained from the
saturated interface widths by imposing the scaling form~(\ref{alpha
definition}) at adjacent values of the system size $L_x$.  Similarly,
the values of $\beta(t)$ are obtained from the transient interface
widths by imposing the scaling form~(\ref{beta definition}) at nearby
times $t$.  We like to remind the reader that simple log-log plots of
$W$ versus $L_x$ and $t$ look typically impressively straight but are
notoriously inaccurate.  The construction of effective exponents, in
the above manner might at first glance look less impressive (the data
appears noisier), but this brings the analysis to a higher level
where the leading corrections to finite-size and finite-time scaling
become visible.

\begin{figure}
{\centering
\resizebox*{0.8\columnwidth}{!}
{\includegraphics{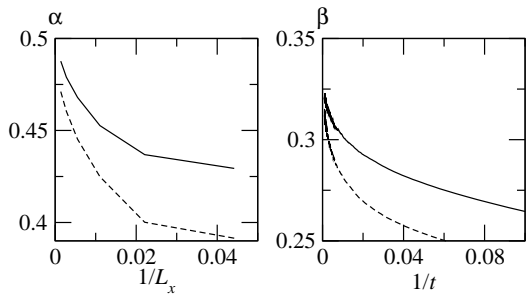}}
\par}
\caption{MC results for the global interface width: left, finite-size
($L_x$) estimates for the saturated surface width exponent $\alpha$;
right, finite-time estimates for the transient interface width
exponent $\beta$ from a flat initial configuration. The solid (dashed)
curves are for continuous (discrete) height model.}
\label{interface scaling exponents}
\end{figure}

The approach to $ L_x\rightarrow\infty $ in Fig.~\ref{interface
scaling exponents} is consistent with the leading correction to
scaling exponent $ y_{{\rm ir}}=-1/2 $ expected from the EW term
$\nabla ^{2}h$ in Eq.~(\ref{KPZ-equation}).  The corrections to FSS
are stronger when the height variables are discrete than when they are
continuous.  This is consistent with the smaller growth rate in the
discrete height interface, and the fact the growth rate is typically
proportional to the nonlinear term $ \lambda $.  On average, more
material is removed during the first deterministic part of the update
process when the surface heights are discrete.

\section{Scaling Properties of 2D avalanches}\label{scaling}

In this section we derive the exact relations between the scaling
properties of the avalanches and 1+1D KPZ interface growth.  However,
in the latter the world sheets are sampled in the correlated manner as
outlined in Sec.~\ref{moti}.

The characteristic feature of SOC is the lack of typical avalanche
length, width, depth, or mass scales.  The probability distributions
follow power laws.  For example, the distribution of avalanche widths
scales as
\begin{equation}
P_{w}\sim w^{-\tau _{w}}
\end{equation}
with scaling exponent $ \tau _{w} $.  Similarly, the avalanche length,
depth, and mass distributions scale as power laws with exponents $
\tau _{l} $, $ \tau _{\delta } $ and $ \tau _{m} $.  We can summarize
this in a metadistribution function $P(l,w,\delta )$; the probability
to find an avalanche of a specific width $w$, length $l$, and depth
$\delta$, obeys the scaling relation
\begin{equation}
P(l,w,\delta )=b^{-\sigma }P(b^{-z}l,b^{-1}w,b^{-\alpha }\delta )
\label{meta}
\end{equation}
with $ b $ an arbitrary scale parameter.  The exponents $ \sigma $, $
z $, and $ \alpha $ are expected to be robust with respect to details
of the dynamic rule, and thus are characteristic of the universality
class to which this avalanche dynamics belongs.  Single parameter
distributions, such as $ P_{w} $, follow by integrating out the other
variables. This implies the following expressions for the $ \tau $
exponents,
\begin{equation}
\label{tau exponent equations}
\tau_{l}=\frac{\sigma-1-\alpha}{z},\,\, \tau_{w}=\sigma-z-\alpha,
\,\, \tau_{\delta}=\frac{\sigma-1-z}{\alpha},
\end{equation}
or inverted,
\begin{equation}
\label{z alpha sigma equation}
z=\frac{\tau_{w}-1}{\tau_{l}-1},\,\, \alpha=\frac{\tau_{w}-1}
{\tau_{\delta }-1},\,\, \sigma=\tau_{w}+z+\alpha.
\end{equation}

Let's presume that the avalanches are compact, i.e., that the inside
and the boundaries of an avalanche are well defined and
distinguishable (unlike in certain fractal structures), and that the
sizes of the holes (unaffected regions) inside the avalanche do not
scale with the avalanche size.  This can be checked visually from
typical simulation configurations, and both assumptions are indeed
satisfied in our dynamics at least qualitatively.  In that case, the
mass of the avalanche must scale as $m\sim l w \delta$, such that the
critical exponent of the distribution of avalanche masses $P_m\sim
m^{-\tau_m}$ obeys the identity
\begin{equation}
\tau_m={ \sigma\over 1+z+\alpha}.
\label{identity1}
\end{equation}

There is one more relation between these critical exponents (leaving
only two independent ones).  The avalanche is initiated by lowering
the bar at the driving edge of the box.  In the stationary state the
average surface profile is invariant, and therefore it shifts down at
the same rate as the lowering bar.  Thus we know how much mass drops
out of the box on average.

To be more precise, during each avalanche event, the height of only
one single boundary site at $ y=0 $ is lowered by, on average, an
amount $ s_{c}/2 $.  For a sandbox of width $ L_{x} $ the boundary row
is lowered by $ s_{c}/2 $ after $ L_{x} $ avalanches.  In the
stationary state, the entire surface matches this lowering speed, such
that the amount of removed sand is on average equal to $ L_{x} L_{y}
s_{c}/2 $.  Therefore, the average mass of each avalanche must be
equal to
\begin{equation}
\label{conservation of average mass}
\left\langle m\right\rangle =\frac{1}{2}s_{c}L_{y}.
\end{equation}
The scaling properties of the mass distribution function tie into this
because
\begin{equation}
\label{conservation of average mass 2}
\left\langle m\right\rangle =\int m'P_{m}(m')dm',
\end{equation}
which can be evaluated using the metadistribution function as
\begin{eqnarray}
\left\langle m\right\rangle  & \sim  &
\int _{0}^{L_{y}}dl\int _{0}^{\infty }dw\int _{0}^{\infty }d\delta \,
lw\delta P(l,w,\delta )\nonumber \\
   & + & m_{L_{y}}\int _{L_{y}}^{\infty }dl\int _{0}^{\infty }dw\int
_{0}^{\infty }d\delta \, P(l,w,\delta ).
\label{mass integrals}
\end{eqnarray}
This equation incorporates finite-size effects.  The box is presumed
to be wide and deep enough, such that the the length $L_{y}$ of the
box (in the direction perpendicular to the driving edge) is the only
limiting finite-size factor.  The first term in the above equation
accounts for all avalanches that fit inside the box and the second
term for the ones that reach the $ L_{y} $ edge, and thus are
prematurely terminated.  The first integral scales as $
L_{y}^{(-\sigma +2+2z+2\alpha )/z} $ for large $ L_{y} $.  The second
term scales with the same power, because the second integral scales as
$ L_{y}^{(-\sigma +1+z+\alpha )/z} $ while the mass factor in front of
it scales as $ m\sim l w \delta \sim L_{y}^{(1+z+\alpha )/z}$.  The
result
\begin{equation}
\label{mass scaling}
\left\langle m\right\rangle \sim L_{y}^{(-\sigma +2+2z+2\alpha )/z},
\end{equation}
when compared to Eq.~(\ref{conservation of average mass}), yields the
exponent identity
\begin{equation}
\sigma =2+z+2\alpha.
\label{identity2}
\end{equation}

The validity of these exponent identities goes well beyond our KPZ
type unloading sandbox.  For example, the EW type directed avalanche
models by Paczuski and Bassler~\cite{Paczuski2000} and Kloster {\it et
al.}~\cite{Kloster2001} obey our Eq.~(\ref{tau exponent equations})
when we substitute for $ z $ and $ \alpha $ the EW values ($ z=2 $, $
\alpha =1/2 $). The scaling exponents of the original Dhar-Ramaswamy
model can be described by the same equations with $ z=2 $, $ \alpha =0
$ as well.

\section{Numerical results for 2D sandbox avalanches}\label{num}

The discussion of the previous section leaves us with two independent
avalanche critical exponents, $\alpha$ and $z$.  The notation
anticipates their identification with the scaling properties of a
rough interface in interface growth.  There, $\alpha$ is the scaling
exponent of the interface width and $z$ the dynamic critical
exponent. Indeed, the interface width relates to the depth of the
avalanche, and time to the the length of the avalanche.  We expect
therefore that $\alpha$ and $z$ take same values as in 1+1D KPZ
growth, $\alpha+z=2$ and $\alpha=1/2$.

We perform MC simulations on the sandbox avalanche model and measure
the avalanche metadistribution function $P(l,w,\delta|L_y)$, see
Eq.~(\ref{meta}).  The sandbox is always taken wide and deep enough
such that the box length $L_{y}$ acts as the only FSS type limiting
factor. We average over $2^{31}$ avalanches.  The reduced
distributions, such as $P_l\sim l^{-\tau_{l}}$, follow from the
metadistribution from, e.g., summation over $w$ and $\delta$.

\begin{figure}
{\centering
\resizebox*{0.8\columnwidth}{!}
{\includegraphics{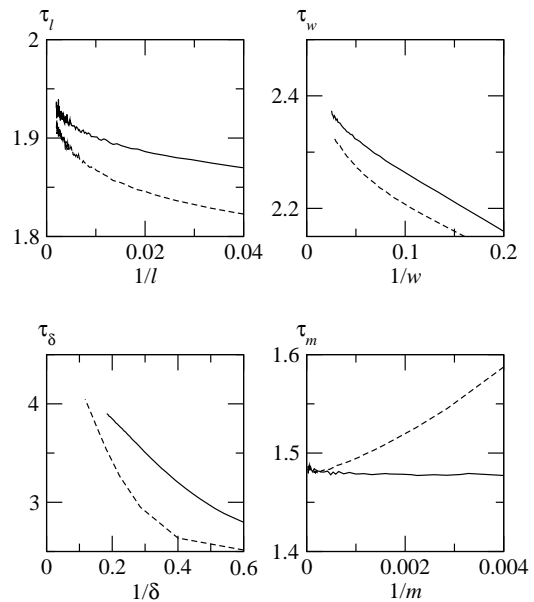}}
\par}
\caption{FSS plots for the $ \tau $ exponents of 2D sandbox model.
The solid (dashed) lines are for continuous (discrete) height model.}
\label{tau exponent plots}
\end{figure}

Figure~\ref{tau exponent plots} shows FSS approximates for the $\tau$
exponents. They are constructed as follows.  Power-law-decaying
objects such as $P_l\sim l^{-\tau_l}$ are almost always subject to
crossover-scaling-type effects, i.e., subdominant additional power-law
terms.  In the language of renormalization theory they originate from
so-called irrelevant scaling fields and also from nonlinear scaling
field effects.  This is well documented in equilibrium critical
phenomena, but most recent nonequilibrium scaling studies ignore this
systematic effect, e.g., by simply making a log-log plot of $P_l$ as
function of $l$ and drawing a least-square-fitting-type straight line
through the data.  Such results show very little statistical noise,
but can give rise to significant systematic errors.  An example of the
importance of corrections to scaling, was the large spread in reported
values of the stationary state roughness exponent $\alpha$ between
various 2D KPZ-type-growth lattice models, which was resolved using a
similar FSS analysis as presented here~\cite{Chin1999}.

In the limit of large $l$ the subdominant additional power-law terms
fade away. So, more weight must be put on the large $l$ part of the
data than on the short $l$ section. However, it is a balancing act,
because at large $l$ the results become noisier, since few avalanches
reach that far.

The total number of avalanches that reach beyond $y$ scales as
\begin{equation}
Q_l(y)=\int_{y}^\infty P_l(l) dl\simeq \frac{A}{\tau_l} ~{y}^{-\tau_l+1},
\end{equation}
if the fraction of avalanches of length $y$ scales as $P_l\simeq A
~y^{-\tau_l}$ (these are only the leading terms).  We construct a $y$
dependent approximate for the exponent $\tau_l$ from the ratio of
these two quantities, as
\begin{equation}
\label{effective tau exponent}
\tau _{l}(y)= \frac{l P_l(y)}{Q_l(y)}
\end{equation}
The results are shown in Fig.~\ref{tau exponent plots}.  (We do the
same for the other distributions.)  Plots such as this are
intrinsically noisier than conventional simple log-log type of plots
of the distributions, but they contain much more information.  The
variation with $y$ reflects the leading corrections to scaling.  The
statistical noise at large $y$ could be suppressed by running the MC
simulation longer.  The simulation time is the only limiting factor.
We used $2^{31}$ avalanches and in that case, $L_y= 512$ is the
optimal box size.

\begin{figure}
{\centering
\resizebox*{0.8\columnwidth}{!}
{\includegraphics{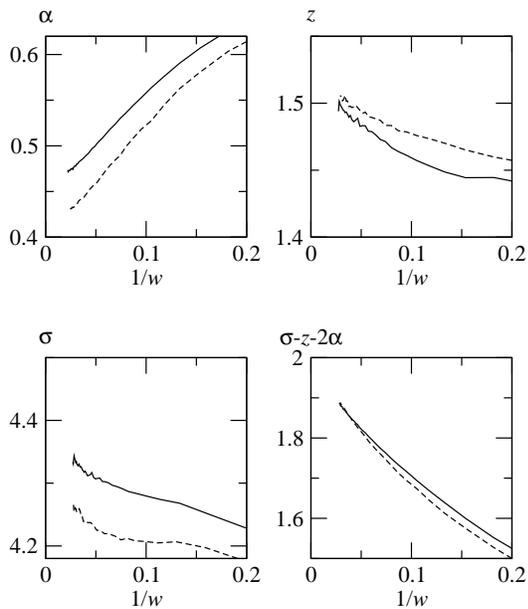}}
\par}
\caption{Effective scaling exponents derived from stationary
avalanche distributions of sandbox systems. The solid (dashed) lines
are for continuous (discrete) height model.}
\label{sigma z alpha plots}
\end{figure}

In Fig.~\ref{sigma z alpha plots}, we replot the same data in terms of
$\alpha$, $z$, and $\sigma$, following Eq.~(\ref{z alpha sigma
equation}) and using the same type of FSS analysis.  From the trend of
the curves, we conclude that $\alpha =0.46\pm 0.01$, $z=1.52\pm 0.02$,
$\sigma =4.43\pm 0.05$, and $\tau_m=1.48\pm0.01$.  This means that the
exponent relations~(\ref{identity1}) and (\ref{identity2}) are
satisfied well within the statistical noise limitations, i.e., within
a few percent.

Surprisingly, the actual values for $z$ and $\alpha$, although close,
differ significantly from the exactly known 1+1D KPZ values,
$\alpha=1/2$ and $z=3/2$.  They deviate more than warranted from
statistical noise alone, and do not converge smoothly if the KPZ
values are correct.  The approximates for $\alpha$ actually undershoot
the KPZ value $\alpha=1/2$, and those for $z$ overshoot $z=3/2$.  This
systematic effect needs to be explained.  It could be that the
exponents differ in a fundamental manner from the conventional KPZ
values, or that we are looking at unusually large and slow corrections
to FSS. The smallness of the deviations makes the latter more likely
(except when this happens to be a continuously varying exponents
scenario).

We will blame the correlated MC averaging feature for this, but it
should be noted that avalanche distributions are intrinsically more
sensitive to FSS effects than global interface features.  Many
avalanches in the ensemble are small compared to the global box size,
and therefore sample and average the KPZ scaling properties over much
smaller lengths and shorter time scales than in a conventional global
interface roughness analysis at a comparable space-time box size.

One option is to push the run button on the computer and out perform
all corrections to FSS. Unfortunately it would require extremely long
MC times, to create large numbers of such large avalanches.  It is
doubtful we would be able to get far enough in a reasonable time span.
Moreover this approach is intellectually unappealing.  We prefer to
search for the origin of the deviations in the exponents.

\section{Avalanche correlated MC runs}\label{corrMC}

The basic premise of our exponent identities is that avalanches are
like any other fluctuation on a 1+1D KPZ type world sheet.  Initially
flat KPZ interfaces (the sand surface next to the driving edge)
roughen in time (moving away from the driving edge) in such a manner
that at (KPZ) time $y$ the stationary state roughness is established
within a length scale $l_x\sim y^{1/z}$.  This defines a so-called
spreading cone.  The avalanches are expected to follow the same
pattern.  However, the avalanche cone seems to spread slightly
faster, since the above avalanche value for $z$ slightly exceeds the
conventional KPZ value, and inside the avalanche the surface seems to
be slightly less rough, since the avalanche value for $\alpha$ is
slightly smaller.

In this and the following section we will establish that this is
caused by correlations with previous avalanches.  The new avalanche
does not run its course on a pristine fresh KPZ interface world sheet
but on an aged one scarred by previous avalanches.

\begin{figure}
{\centering
\resizebox*{0.8\columnwidth}{!}
{\includegraphics{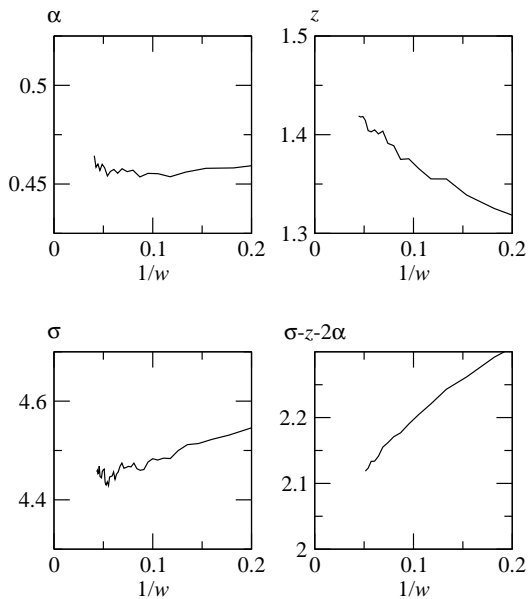}}
\par}
\caption{Effective scaling exponents derived from the distributions
of first avalanches on fresh sandbox surface for the continuous height
model}
\label{fresh azs exponents}
\end{figure}

There are two obvious tests to address the effects of these scars.
The first one is to determine the avalanche distributions for only the
first avalanche on a fresh KPZ world sheet (the initial condition),
i.e., to refresh the entire surface completely after each avalanche.
The results are shown in Fig.~\ref{fresh azs exponents}.  The
first-avalanches likely follow normal KPZ exponents: $z$ converges now
smoothly towards $z=3/2$; while the FSS approximates for $\alpha$,
although still too small, start to turn towards $\alpha=1/2$, and do
not cross that value anymore.  It should be noted that the FSS
corrections are expected to be larger, and that the data is noisier
than in Fig.~\ref{sigma z alpha plots}, because although we ran the
same number of avalanches ($2^{31}$), the fraction of large avalanches
is smaller, leading to smaller and noisier amplitudes in the power-law
tails of the distributions.

The second test of the role of the scars is to measure the global
interface roughness for avalanche type correlated MC runs instead of
completely refreshing MC runs.  The upper panel of Fig.~\ref{comparing
square width} shows the global interface width $W^2$ as function of
time for several $L_x$s.  The drawn lines correspond to avalanche
correlated MC runs and the dashed line to conventional uncorrelated MC
averaging.  The drawn lines have bumps, i.e., the avalanche correlated
runs lead to rougher interfaces at intermediate times.

\begin{figure}
{\centering
\resizebox*{0.8\columnwidth}{!}
{\includegraphics{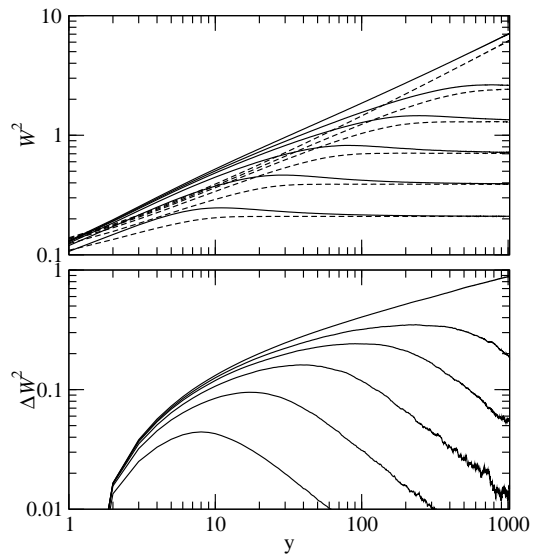}}
\par}
\caption{Upper panel: square interface width for stationary sandbox
surface (solid lines) comparing with fresh surface (dashed lines);
Lower panel: the difference between the two. From bottom up, the
corresponding system sizes, $ L_{x} $, in the transverse (x) direction
are 8, 16, 32, 64, 128 and $ \infty $.}
\label{comparing square width}
\end{figure}

\begin{figure}
{\centering
\resizebox*{0.8\columnwidth}{!}
{\includegraphics{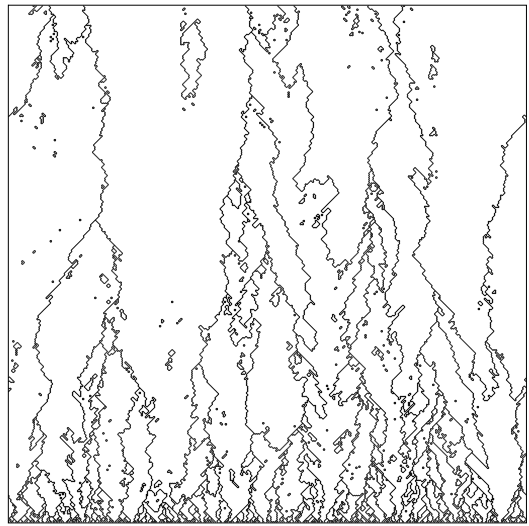}}
\par}
\caption{A typical configuration of the scars on the sandbox created
by the avalanches. The driving edge is located at the bottom of the
graph while avalanches propagate upward in the $y$ (or $t$)
direction. The system sizes are $L_x = 256$ and $L_y=512$.}
\label{avalanche scars}
\end{figure}

This enhanced interface roughness is caused by the scars left by
earlier avalanches.  The scars vanish at very large $y$ because
avalanches reaching that far span the entire system in the
$x$-direction.  Figure~\ref{avalanche scars} shows a typical
configuration of scars.  The lines are the traces of previous
avalanches, i.e., their edges.  Latter avalanches wipe them out
partially.

For finite system sizes, the stationary state interface width follows
from the plateaus at large times.  There the avalanche correlated and
uncorrelated MC curves coincide.  This is to be expected, because the
large avalanches that span the entire system (in the $x$ direction at
large $y$) occur at regular MC time intervals, such that the large $y$
part of the surface (i.e., the stationary state of the growth process)
is completely refreshed periodically and therefore sampled effectively
like in uncorrelated MC runs.  As a result, the roughness exponent
$\alpha$, defined by Eq.~(\ref{alpha definition}), is the same for the
both cases.

Most avalanches do not extend into that large $y$ part of the surface.
They terminate in the scarred part of the surface.  Therefore, we
define an alternative roughness exponent $\alpha^*$, associated with
the scaling of the bumps, in terms of the maximized width
\begin{equation}
\label{alpha from maximized width}
  W^* \equiv \max_y
  W(L_x,y) \sim L_x^{\alpha^*}
\end{equation}
more relevant for the avalanche scaling properties.  Note that for
uncorrelated MC runs, $\alpha^* = \alpha$, since the interface width
increases monotonically in time.

The conventional method for measuring the exponent $\beta$, involves
the slope at times $y<L_x^z$, and thus is sensitive to the bumps in $W$
as well.  The results are shown in Fig.~\ref{interface exponents
corr-MC}. Compared to those in Fig.~\ref{interface scaling exponents},
they clearly converge less smoothly, with larger corrections to
scaling and we should wonder if they converge to the conventional
exact KPZ values, $\alpha=1/2$ and $\beta=1/3$, at all.

\begin{figure}
{\centering
\resizebox*{0.8\columnwidth}{!}
{\includegraphics{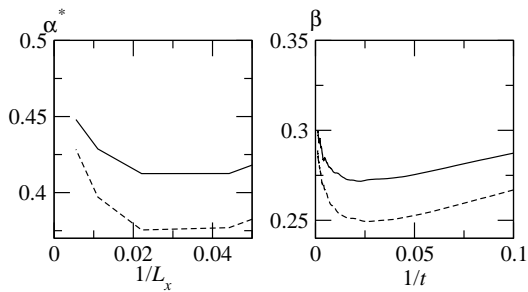}}
\par}
\caption{Finite-size approximates of the scaling exponents for
stationary surface of sandbox (or correlated MC runs for the interface
model) with $\alpha^*$ defined by Eq.~(\ref{alpha from maximized
width}) and $\beta$ by Eq.~(\ref{beta definition}). The solid (dashed)
curves are for the continuous (discrete) height model.}
\label{interface exponents corr-MC}
\end{figure}

In the lower panel of Fig.~\ref{comparing square width} we plot
$\Delta W^2$ as function of time, the difference between the squared
widths of avalanched correlated MC runs (the drawn lines in the upper
panel) and completely uncorrelated MC runs (the dashed lines in the
upper panel).  For infinite system size, $\Delta W^2$ scales as
$\Delta W^2\sim y^{s}$ with an exponent that numerically is very close
to $s\simeq {1/3}$.  Since the width itself scales as $W^2\sim
y^{2/3}$, it follows that the bumps in the width curves are a
transient FSS effect.

This settles our basic issue at the numerical level; the avalanche
correlated nature of the MC runs does not change the interface scaling
exponents, but only gives rise to slow corrections to FSS.  In the
next two sections we will identify these corrections to scaling with
the scars on the surface left behind by previous avalanches.

We start this analysis here by casting the deviations into the
framework of corrections to scaling from a so-called irrelevant
operator in the sense of renormalization theory.  Let $O_{\rm sc}(x)$
be that irrelevant operator and $u$ be its scaling field.  This mounts
to presuming that the avalanche correlation between MC runs can be
represented effectively by adding to the KPZ Langevin
equation~(\ref{KPZ-equation}), a term $u O_{\rm sc}(x)$.  We will have
to determine below how $O_{\rm sc}(x)$ is related to the density of
scars on the interface space-time world sheet left by previous
avalanches.  According to scaling theory, the presence of such a term
to the Langevin equation leads to corrections to scaling in the
interface width as
\begin{equation}
W^2(L_x, y, u) = b^{2\alpha} W(b^{-1}L_x, b^{-z}y, b^{y_{\rm sc}} u),
\end{equation}
i.e., in the infinite-size limit, $L_x\to \infty$, to
\begin{equation}
W^2(y, u) = y^{2\alpha /z} S(y^{y_{\rm sc}/z} u),
\end{equation}
and by expanding the scaling function $S$, while assuming that $y_{\rm
sc}<0$, such that $u=0$ is a stable fixed point, and the argument
$y^{y_{\rm sc}/z} u$ is a small parameter, to
\begin{equation}
\label{irrelevant}
W^2(y, u) = y^{2\alpha/z}\left[ S(0) +  y^{y_{\rm sc}/z} u
S^\prime(0) + \cdots\right].
\end{equation}
The critical exponent $y_{\rm sc}$ of this irrelevant scaling field
must take the value $y_{\rm sc}=-\alpha$ to account for the $\delta
W^2\sim y^{1/3}$ corrections in the interface width we found above.
Moreover the operator must scale as
\begin{equation}
O_{\rm sc}(x) \sim b^{-x_{\rm sc}}
\end{equation}
with critical dimension $x_{\rm sc}= z$, since the KPZ
equation~(\ref{KPZ-equation}), implies that the terms $u O_{\rm
sc}(x)$ and $\partial h/\partial t$ must scale alike.  In the
following two sections we will trace down the geometric identity of
this mysterious operator $O_{\rm sc}$, starting with the 1D version of
the model.

\section{Surface rounding in the 1D unloading sandbox}\label{1Dedge}

The 1D version of the unloading sandbox shows the same type of
differences between uncorrelated and avalanche-type correlated MC runs
as the 2D version.  We determined numerically the difference between
the interface width for avalanche-correlated and uncorrelated MC runs,
and found that it diverges as a power law $\delta W^2\sim y^{1/2}$,
with an exponent which is again (like in 2D) half the size of that for
$W^2\sim y$ itself.  According the corrections to scaling
formalism~(\ref{irrelevant}), the scaling dimension of $O_{\rm sc}$
must therefore be equal to $x_{\rm sc}=z$, just as in 2D.

The underlying interface dynamics becomes a zero dimensional growth
model, i.e., a simple random walk in the $h$ direction with a nonzero
drift velocity to account for the net tilt of the surface.  The
exponents of the various avalanche distribution functions must obey
the same type of relations as in Sec.~\ref{scaling}
\begin{equation}
\tau_l=\frac{\sigma-\alpha}{z},
\quad
\tau_\delta=\frac{\sigma-z}{\alpha},
\quad
\tau_m=\frac{\sigma}{\alpha+z},
\end{equation}
  and
\begin{equation}
\sigma=z+2\alpha
\end{equation}
Without loss of generality we can set $\alpha=1$ (measure all lengths
in terms of $\delta$).  These identities are satisfied exactly, and
the exponents are the same for uncorrelated and avalanche correlated
runs.  From the interface dynamics perspective, a single directed
random walker, the diffusion equation character of the dynamics
implies that $z=2\alpha=2$.  The values of all the other exponents
follow from this, and are consistent with their values from the
avalanche perspective.  There, we are dealing with the statistics of
merging random walkers.  The number of walkers at a given ``time'' $y$
is equal to the number of avalanches of a length $l$ equal or larger
than $y$ in the ensemble of MC runs.  The density of the walkers
decays as $\rho (y)\sim y^{-1/2}$ \cite{Hinrichsen1997}, such that the
distribution of avalanche lengths obeys the form
\begin{equation}
P_{l}(l)=\left[ -\frac{\partial }{\partial y}\rho (y)\right]
_{y=l}\sim l^{-3/2},
\end{equation}
and therefore that $ \tau_{l}=3/2 $.  The depth of the avalanche
follows from the maximum separation between two subsequent walkers,
and scales as $\delta \sim l^{1/2} $, i.e., $\alpha/z=1/2$.  The mass
scales as $ m\sim l \delta \sim l^{3/2} $, i.e., $(\alpha+z)/z= 3/2$
and $\tau_m = 4/3$.

This can be compared directly with the exponents of other 1D sandpile
models, e.g., with results by Paczuski and
Boettcher~\cite{Paczuski1996} on the so-called Oslo sandpile model
where $\tau \equiv \tau_m \approx 1.55$ and $D \equiv (\alpha + z)/z
\approx 2.23$.

Let's turn our attention now to the central issue, the difference
between uncorrelated versus avalanche-correlated MC runs.  Adding a
term like $u O_{\rm sc}$ to the diffusion equation of motion creates a
correction to the drift velocity of the random walk.  This suggests we
can identify the geometric meaning of $O_{\rm sc}$ directly by
studying the deviations of the slope near the driving edge of the
surface from its asymptotic value.

The average surface slope does not show any deviations (near the
driving edge) from $s_c/2$ when we run the dynamics as a conventional
random walk, which amounts to ``completely refreshing'' the surface
after each MC run (uncorrelated MC runs).  The avalanche-correlated
runs do show a rounding of the surface near the driving edge
\begin{equation}
s(y)\simeq A y^{-\kappa} + \frac{1}{2} s_c
\end{equation}
The numerical results for the exponent yield $\kappa= 0.98\pm 0.03$,
in accordance with $x_{\rm sc}=z$ from the interface width since
$\kappa= x_{\rm sc}/z$ and $z=2$ for random walks.

This rounding originates from the distribution of termination points
of the avalanches.  A new random walk starts below the previous one
and propagates until it meets the previous trajectory and terminates.
The avalanche is the space between the trajectory of that new random
walk and the already existent surface.  The amount of rounding of the
slope near the driving edge is proportional to the distribution
$\rho(y)$ of merging points on the surface.  Those are the scars from
previous avalanches.  Each random walker by itself does not contribute
to the rounding, i.e., on average its walk has constant slope $s_c/2$
independent of time.  However the merging process truncates each walk
and does so in an upwards biased fashion.  Each merging event causes
the surface to drift upwards by a certain amount ($s_c/2$, in average,
for the discrete $h$ version).  Therefore the rounding of the surface
is proportional to $\rho(y)$.

\begin{figure}
{\centering
\resizebox*{0.8\columnwidth}{!}
{\includegraphics{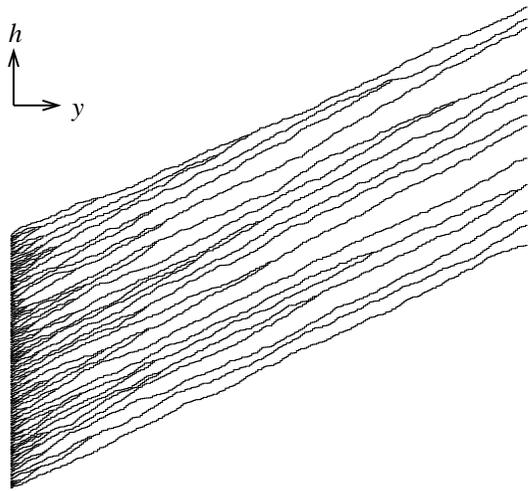}}
\par}
\caption{Traces of stable sand surface over 256 avalanches for 1D
sandbox model with $L_{y} = 256$. The system is driven from the left
at $y = 0$.}
\label{traces of 1D sand surface}
\end{figure}

The entire process and the set of subsequent stable sand surfaces
(Fig.~\ref{traces of 1D sand surface}) is therefore equivalent to a
system of merging random walkers obeying the rule $A+A\rightarrow A$.
That type of dynamics has received extensive attention recently and
its various scaling properties are known
exactly~\cite{Hinrichsen1997}.  There is little doubt that our 1D
unloading sandbox is exactly soluble, using absorbing wall type random
walk mathematics~\cite{group}.  However, we will refrain from pursuing
this path in this paper.

The critical dimension of $O_{\rm sc}\sim \rho(y)$ can be estimated
(for intuition building purposes) as follows.  After adding a term
$u O_{\rm sc}$ to the KPZ equation we should also write down an
equation of motion for $O_{\rm sc}$ itself, to close the equations.
The latter is not trivial, because the scars on the surface build up
slowly in time, such that that the equation of motion for $O_{\rm sc}$
is highly nonlocal.  On the other hand, the linear nature of the
diffusion equation allows one to be somewhat frivolous with the order
in which averages are taken, (without losing the essential physics,
nor even the correct critical exponents).

Let $\rho_{\tilde t}(y)$ be the endpoint distribution after $\tilde t$
avalanches (MC time steps).  During the last MC time step, one
avalanche runs through the system.  It refreshes the entire surface
before its termination point $y=l_{\tilde t}$, such that $\rho_{\tilde
t}$ at site $y$ does not change if the avalanche terminates before
$y$; $\rho_{\tilde t}(y)=1$ if it terminates at $y$; and $\rho_{\tilde
t}(y)=0$ if it extends beyond $y$:
\begin{equation}
\frac{\partial \rho_{\tilde t}(y) }{\partial {\tilde t}} = P_l(y) -
\rho_{\tilde t}(y) \int_y^\infty P_l(l) dl
\end{equation}
with $P_l(l)$ the probability that the avalanche terminates at
distance $l$ from the driving edge.  The stationary state endpoint
profile therefore takes the form
\begin{eqnarray}
\rho (y) = \frac{P_{l}(y)}{\int _{y}^{\infty}P_{l}(l)dl},
\label{endpoint integration}
\end{eqnarray}
and $ P_{l}(l) \sim l^{-\tau _{l}} $ yields
\begin{equation}
\label{1D decay of slope change}
\rho (y)=\frac{1}{\tau _{l}-1}y^{-1}.
\end{equation}
In other words, the surface curvature scales as $ \Delta s\sim
y^{-x_{\rm sc}/z }$ with $x_{\rm sc}=z$, in agreement with the above
results.  Interestingly, this result is independent of the actual
value of the scaling exponent $\tau_l$, provided that $ \tau _{l}>1 $,
which has to be true for $ P_{l} $ to be normalizable.

In conclusion, in 1D we identified the crossover scaling operator with
the density of avalanche endpoints. These represent indeed the scars
on the surface, the memory of previous avalanches.

\section{Avalanche rounding near the driving edge in 2D}\label{2Dedge}

As in the 1D model, the surface slope is modified by the iterated
avalanche process.  However, unlike in 1D, the average slope near the
edge is not constant already in conventional interface dynamics (where
the entire surface is being refreshed during each MC run).  The
surface slope is related to the growth rate of the underlying
interface, and the rounding of the slope near the driving edge
represents the transient growth rate of the KPZ interface from the
initial configuration, e.g., a flat one:
\begin{equation}
s_{\rm f}(y)\simeq v_{0}+cy^{-\kappa_{\rm f}}
\label{free-s}
\end{equation}
with $y$ playing the role of time and the subscript, ${\rm f}$,
denoting that the entire surface is refreshed.  By direct numerical
simulation of uncorrelated interface dynamics we find $\kappa_{\rm f}
\approx 0.7$ (the left panel of Fig.~\ref{scaling exponents of slope
change plot}). This is consistent with conventional KPZ scaling and
power counting
\begin{equation}
s\sim h/y\sim y^{\alpha/z-1}\sim y^{-2/3},
\end{equation}
suggesting $\kappa_{\rm f} =2/3$.

\begin{figure}
{\centering
\resizebox*{0.8\columnwidth}{!}
{\includegraphics{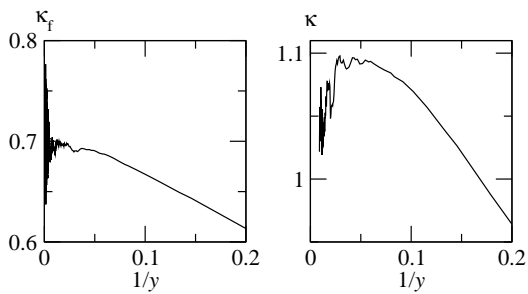}}
\par}
\caption{Scaling exponent for boundary correction to the local slope
of fresh 2D sandbox surface (or, in the interface language, transient
growth rate from a flat interface), $s_{\rm f} (y) - s_{\rm f}
(\infty)\sim y^{-\kappa _{\rm f}}$, and its correction due to the
iterated avalanche process, $\Delta s = s(y) - s_{\rm f} (y)\sim
y^{-\kappa }$.}
\label{scaling exponents of slope change plot}
\end{figure}

We evaluate the surface slope profile $s(y)$ in avalanche correlated
dynamics MC runs, in terms of the difference with respect to the
uncorrelated case,
\begin{equation}
\Delta s(y)=s(y)-s_{\rm f}(y) \sim y^{\kappa}
\end{equation}
The FSS analysis for the exponent $\kappa$ (the right panel of
Fig.~\ref{scaling exponents of slope change plot}) yields $\kappa
=1.05\pm 0.07$.  This is in agreement with $x_{\rm sc}=z$ and
$\kappa=x_{\rm sc}/z$ implied by the corrections to scaling
formalism~(\ref{irrelevant}).

Inside the bulk of an avalanche the interface is fully refreshed, and
scales as in uncorrelated KPZ dynamics.  At the avalanche boundaries,
the slope of the surface is biased upwards, because of the merging
with previous MC runs (which are on average shifted upwards by an
amount $s_c/2L_x$ each time an avalanche is triggered).  This means
that the $\Delta s$ is proportional to the density of scars in the
surface.  In 1D, the scars are point-like objects, the endpoints of
the avalanches; but in 2D the avalanche boundaries are line objects.
This nonscalar aspect makes that most line-segment contributions, when
integrated along the boundaries of an avalanche, cancel out against
each other.

\begin{figure}
{\centering
\resizebox*{0.8\columnwidth}{!}
{\includegraphics{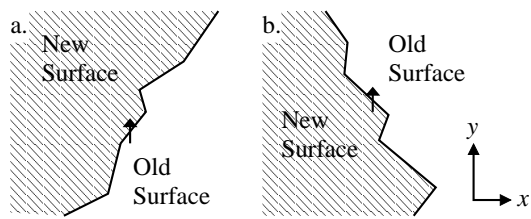}}
\par}
\caption{Two possible cases at a boundary of an avalanche cluster
(the shaded area): a. avalanche expands; b. avalanche shrinks. The
local slopes along the arrow marks is reduced in a. while increased in
b.}
\label{edge of avalanche}
\end{figure}

To be more precise, $s(y)$ represents only the component of the slope
in the $y$-direction, and the magnitude of those jumps depends on the
local angle $\theta$ the boundary makes with the $y$-axis.  This is an
odd function, $\Delta (\theta)=-\Delta(-\theta)$, as illustrated in
Fig.~\ref{edge of avalanche}.  The slope change is negative when the
avalanche opens up and positive when it narrows down.  The latter also
implies that $\Delta (\theta)$ has opposite sign for the left and
right boundary of each avalanche.  Notice that, while in the lattice
model $\theta$ takes only two discrete values, it renormalizes to a
continuous variable at larger length scales.

Let's estimate the change in surface slope due to these scars in the
same spirit as we did successfully in 1D.  Consider one specific
surface, and let $s_{\tilde t}(y)$ be the surface slope in a slice of
the surface at distance $y$ from the driving edge, averaged over all
$x$, after ${\tilde t}$ avalanches (MC time ${\tilde t}$).  The last
avalanche changes this as follows.  Let $w_{\tilde t}(y')$ be the
width of this avalanche, which terminates at $y=l_{\tilde t}$, in
slice $y'$.  The inside area of the avalanche is completely refreshed
and therefore has the same average slope $s_{\rm f}(y)$ as in ordinary
KPZ dynamics (totally refreshed subsequent world sheets).  This leads
to the following equation of motion,
\begin{equation}
\frac{\partial s_{\tilde t}(y)}{\partial {\tilde t}}= [\Delta(\theta_L)
-\Delta(\theta_R)]  + w_{\tilde t}(y)[s_{\rm f}(y)-s_{\tilde t}(y)]
\label{bulk slope change}
\end{equation}
The first term on the right hand side represents the creation of the
two new avalanche edges, and the second term represents the refreshed
surface inside the new avalanche.  Note that $\partial s_{\tilde
t}(y)/\partial {\tilde t}=0$ when this latest avalanche does not reach
slice $y$, and that this is automatically taken care of because in
that case $\theta_L=\theta_R=0$ and $\Delta(0)=0$, while $w_{\tilde
t}(y)=0$ for $y>l_{\tilde t}$.  In the stationary state, after
averaging over all possible avalanches, Eq.~(\ref{bulk slope change})
leads to
\begin{equation}
\overline{w_{\tilde t}(y)\left[s_{\rm f}(y)-s_{\tilde t}(y)\right]} =
\overline{\Delta(\theta_L) -\Delta(\theta_R)}
\end{equation}
Next, we perform an heuristic coarse-graining renormalization-type
transformation.  At large length scales, the average angle $\theta$
remains small, such that the right hand side can approximated as
\begin{equation}
\overline{\Delta(\theta_L) -\Delta(\theta_R)}
\simeq
a~ \overline{\theta_L-\theta_R}
\simeq a~ \frac{\partial \overline{w_{\tilde t}(y)}}{\partial y}
\end{equation}
Finally, we presume that in the stationary state it is not too bad to
treat the KPZ height fluctuations deep inside the bulk of an avalanche
and those near its edge as decoupled (at least in lowest order) such
that
\begin{equation}
\Delta s(y) = \overline{s_{\rm f}(y)-s_{\tilde t}(y)} =
a~\frac{\partial}{\partial y} \log(\overline{w_{\tilde t}(y)}).
\end{equation}
This yields $\Delta s(y)\sim y^{-1}$, exactly the power-law decay we
are looking for, and consistent with all the above numerical results.

The only requirement for the latter is that $\overline{w_{\tilde
t}(y)}\sim y^{-\xi}$ decays as a power law.  Again, like in
Eq.~(\ref{1D decay of slope change}) for 1D, the value the critical
exponent $\xi$ does not matter.  $\overline{w_{\tilde t}(y)}$ is
equal to the average avalanche width in slice $y$ averaged over all
avalanches.  It is reasonable to expect, and we confirmed numerically,
that this quantity scales with the same exponent as the average width
of all avalanches longer than $y$, i.e., as
\begin{equation}
\int_y^\infty w(l) P(l) dl \sim y^{1/z-\tau_l+1}
\end{equation}
which yields $\xi\simeq 1/3$.

We are now ready to represent the crossover scaling operator $O_{\rm
sc}(x) $ in terms of the scars on the surface. Consider time slice
$y$. $O_{\rm sc}(x)=0$ when no scar line runs through site $x$, and
otherwise is proportional to the angle the scar line makes with
respect to the $y$-axis.  However the sign also flips depending on
whether this represents a left or right boundary of the original
avalanche.  The latter can be denoted by an arrow along the avalanche
scar line.  Alternatively, we can associate an {\it age}-field
$g(x,y)$ to the entire surface, representing the age of the surface
segments (how many MC time steps ago site $x$ was updated),
\begin{equation}
\label{scar-operator}
O_{\rm sc}\sim \frac{\hat e_y \cdot \nabla g}{|\nabla g|}
\end{equation}
with $\hat e_y$ a unit vector in the $y$-direction.  The denominator
arises because the magnitude of the age jump across the scar line
$|\nabla g|$ does not play a role.

\section{Summary}\label{sum}

In this paper, we studied a directed avalanche model inspired by the
unloading of a sandbox by means of a slowly lowering wall, and the
wish to setup an avalanche dynamic rule belonging to the same
universality class as KPZ type interface growth. The 2D sand surface
represents the world sheet of the 1+1D growing interface.

The scaling exponents of the avalanche distributions are directly
related to the dynamical and stationary state roughness exponents $z$
and $\alpha$ of KPZ growth in 1+1D, Eq.~(\ref{tau exponent
equations}).  However, we encounter one crucial difference. From the
avalanche perspective the conventional uncorrelated MC runs correspond
to completely refreshing the surface, i.e., an ensemble average over
all possible initial conditions, without ever running an avalanche.
From the KPZ perspective, the avalanche dynamics represents an unusual
MC ensemble averaging procedure where subsequent interface world
sheets only differ inside the single avalanche.  This
avalanche-correlated-type averaging enhances the interface roughness
at time scales $y<L_x^z$, due to the scars of previous avalanches.  It
required a careful study, combining numerical and analytical tools,
presented in the second half of this paper, to establish that these
scars give rise only to larger than usual corrections to scaling and
not to fundamentally different values of the global roughness scaling
exponents $z$ and $\alpha$.

The effect of the scars can be represented by introducing an
additional age field $g(x,y)$ to the height variables $h(x,y)$, that
keeps track of how many MC runs ago site $(x,y)$ participated in an
avalanche.  This age-field couples into the KPZ
equation~(\ref{KPZ-equation}) as an additional term of the form $u
O_{\rm sc}$. The operator $O_{\rm sc}$ is proportional to the angle a
scar makes with respect to the time-axis, and can be expressed in
terms of the age field as shown in Eq.~(\ref{scar-operator}).  We
establish that the coupling of this age field to the KPZ equation is
irrelevant in the sense of renormalization theory, both numerically
and by writing down approximate equations of motion for $u O_{\rm
sc}$.  The scaling field $u$ renormalizes with exponent $y_{\rm
sc}=-\alpha$ and $O_{\rm sc}$ scales with critical dimension $x_{\rm
sc}=-z$.

We believe that the results of our work presented here can be
generalized to most ``Markovian'' avalanche dynamic systems with local
row-by-row type toppling rules, and that this is a promising route to
improve our understanding of the scaling properties of avalanche
dynamics in general.

This research is supported by the National Science Foundation under
grant DMR-9985806.

\end{document}